\documentclass{aa}  

\usepackage{graphicx}
\usepackage[switch]{lineno}
\usepackage{txfonts}
\usepackage{hyperref}
\newcommand{\fgbm}{\textit{Fermi}/GBM\xspace}
\newcommand{\fermilat}{\textit{Fermi}/LAT\xspace}
\newcommand {\swift} {\textsl{Swift}\xspace}
\newcommand {\Fermi} {\textsl{Fermi}\xspace}

\newcommand {\kwind} {Konus/\textsl{Wind}\xspace}

\newcommand {\integral} {INTEGRAL\xspace}

\begin{document}

   \title{From Rare Events to a Population: Discovering Overlooked Extragalactic Magnetar Giant Flare Candidates in Archival Fermi Gamma-ray Burst Monitor Data}
   \titlerunning{Archival GBM MGF Candidates}

   % \subtitle{I. Overviewing the $\kappa$-mechanism}

   \author{Aaron C. Trigg
          \inst{1}
          \and Eric Burns \inst{2}
          \and Michela Negro \inst{2}
          \and Suman Bala \inst{3}
          \and P.N. Bhat \inst{4}
          \and William H. Cleveland\inst{3}
          \and Dmitry D. Frederiks\inst{5}
          \and Adam Goldstein\inst{3}
          \and Boyan A. Hristov\inst{4}
          \and Daniel Kocevski \inst{6}
          \and Niccol\`{o} Di Lalla \inst{7}
          \and Stephen Lesage \inst{4}$^,$ \inst{8}
          \and Bagrat Mailyan \inst{9}
          \and Eliza Neights \inst{10}$^,$ \inst{11}
          \and Nicola Omodei \inst{7}
          \and Oliver J. Roberts\inst{3}
          \and Lorenzo Scotton\inst{4}
          \and Dmitry S. Svinkin\inst{5}
          \and Joshua Wood\inst{6}
            }
    \institute{NASA Postdoctoral Program Fellow, NASA Marshall Space Flight Center, Huntsville, AL, 35812, USA \\
            \email{aaron.c.trigg@nasa.gov}
        \and Department of Physics \& Astronomy, Louisiana State University, Baton Rouge, LA 70803, USA
        \and Science and Technology Institute, Universities Space Research Association, Huntsville, AL 35805, USA
        \and Center for Space Plasma and Aeronomic Research, University of Alabama in Huntsville, Huntsville, AL 35899, USA
        \and Ioffe Institute, 26 Politekhnicheskaya, St. Petersburg, 194021, Russia
        \and  ST12 Astrophysics Branch, NASA Marshall Space Flight Center, Huntsville, AL 35812, USA
        \and W. W. Hansen Experimental Physics Laboratory, Kavli Institute for Particle Astrophysics and Cosmology, Department of Physics and SLAC National Accelerator Laboratory, Stanford University, Stanford, CA 94305, USA
        \and Department of Space Science, University of Alabama in Huntsville, 320 Sparkman Drive, Huntsville, AL 35899, USA
        \and Department of Aerospace, Physics and Space Sciences, Florida Institute of Technology, Melbourne, FL 32901, USA
        \and Department of Physics, The George Washington University, 725 21st St NW, Washington, DC 20052, USA
        \and Astrophysics Science Division, NASA Goddard Space Flight Center, 8800 Greenbelt Road, Greenbelt, MD 20771, USA
            }

   \date{}

% \abstract{}{}{}{}{} 
% 5 {} token are mandatory
 
  \abstract
    {Magnetar giant flares are rare, extremely bright bursts of gamma-rays from highly magnetized neutron stars. These events are challenging to identify because, at extragalactic distances, they can appear similar to other astrophysical phenomena. Only a handful have been confidently identified to date, limiting our understanding of their origin and physical properties. This study focuses on expanding the sample of known events and enabling a more detailed characterization of their observational features and intrinsic properties, while introducing significant improvements in the methods used to identify and analyze them. When applied to archival data from the Gamma-ray Burst Monitor (GBM) on the \Fermi Gamma-ray Space Telescope, this approach added four previously unidentified events the known sample, expanding the total to 13 MGFs. This demonstrates both the effectiveness of the method and the likelihood that additional magnetar giant flares remain hidden in existing gamma-ray burst catalogs. We utilize this expanded sample to gain a deeper understanding of the broader population of magnetar giant flares. We develop a statistical modeling framework that combines previously considered data with modern observations from \fgbm. The model accounts for instrumental sensitivity and the expected diversity in event characteristics. We infer a volumetric rate of events above $1.2\times10^{44}\,\rm{erg}$ of $R_{MGF}=5.5^{+4.5}_{-2.7}\times10^5\rm{Gpc^{-3}yr^{-1}}$. The results show that individual magnetars must produce multiple flares throughout their lifetimes, reinforcing the idea that these are recurring phenomena rather than singular explosive events. Expanding the sample of known magnetar giant flares improves our understanding of magnetars and their role in other astrophysical phenomena, including possible links to fast radio bursts, gravitational waves, and the creation of heavy elements in extreme astrophysical environments.}

    % \titlerunning
   \maketitle
%
%________________________________________________________________

\section{Introduction}

    A core collapse supernova (CCSN) marks the catastrophic end of a massive star, resulting in the formation of either a black hole or a neutron star (NS). In some cases these NSs are characterized by the strongest persistent magnetic fields in the cosmos, exceeding $10^{14}\,\rm{G}$ \citep{1984Ap&SS.107..191U,duncan1992formation,Paczynski:1992zz,ThompsonDuncan1995,ThompsonDuncan1996,kaspi_magnetars_2017}. These NSs are classified as magnetars. The densities and field strengths of these compact remnants enable studies of physics inaccessible in laboratory environments. More than two dozen magnetars have been identified in star-forming regions of the Milky Way \citep{Gaensler2004AdSpR..33..645G} and Large Magellanic Cloud \citep[LMC;][]{Olausen_2014}, where they produce a variety of high-energy transient emissions. The diverse range of phenomena includes sub-second bursts, burst "storms" made up of hundreds to thousands of bursts over several minutes, and the most extreme example of magnetar emission, the magnetar giant flare \citep[MGF;][]{kaspi_magnetars_2017,Negro_2024_frontiers}. 
    
    Magnetar giant flares initially display a brief, millisecond-long gamma-ray peak that is spectrally harder than the typical magnetar emission, with isotropic-equivalent energies $E_{\rm iso}\sim\,10^{44}-10^{46}\,\rm{erg}$. The initial hard, high-energy peak then transitions to a weaker, softer tail that last several minutes. This tail displays a periodic signal due to modulation by the spin period of the source NS \citep{hurley1999giant,Palmer+05}. Since the first observation of a MGF, localized to the magnetar SGR 0526–66 in the LMC on 3 March 1979 \citep{mazets79Natur}, two additional  MGFs have been associated with the magnetars SGR\,1900+14 \citep{hurley1999giant,Feroci1999} and SGR\,1806-20 \citep{Palmer+05,Frederiks2007AstL...33....1F} in the Milky Way. The initial peaks for all three events exhibited extremely high flux rates that saturated the detectors on nearly all observing instruments.

    However, at extragalactic distances, the rotationally modulated tail of an MGF is too faint to be detected by current instruments without rapid identification and re-pointing \citep{Negro_2024_frontiers, Trigg2025A&A...694A.323T}, and has yet to be observed. Therefore, a distant event lacking such a tail would appear similar to, and potentially misidentified as, a cosmological short gamma-ray burst \citep[sGRB;][]{Mazets1982Ap&SS..84..173M,Duncan2001AIPC..586..495D, Hurley+05,Palmer+05,Hurley2011}, a phenomenological class defined by transient flashes of high-energy photons in the 10\,keV–10\,MeV range that last less than $\sim$2 seconds believed to originate from mergers of compact object binaries such as NS–NS or NS–black hole (BH) systems \citep{kouveliotou2012gamma, Kouveliotou1993, Eichler1989, Fong2015ApJ...815..102F, Abbott2017a, Abbott2017b, Goldstein2017}. This limitation requires alternative methods for identifying MGFs that originate outside of the Milky Way: localizing sGRBs to nearby star-forming galaxies.

    Six extragalactic bursts have been classified as MGFs, relying primarily on their spatial coincidence with star-forming galaxies, and supported by their temporal and spectral properties. Five were localized by the Interplanetary Network of gamma-ray satellites \citep[IPN;][]{Hurley2013IPN} and one by the International Gamma-ray Astrophysics Laboratory (\integral), enabling associations with nearby star-forming galaxies. These candidates are GRB\,051103, associated with M81 \citep{ofek2006short,Frederiks_2007AstL...33...19F,Hurley_2010new}, GRB\,070201, associated with the galaxy M31 \citep{Mazets_2008,Ofek_2008}, GRB\,070222, associated with M83 \citep{2021Burns}, GRB\,200415A and GRB\.180128A, both localized to NGC\,253  \citep{svinkin2021bright,Roberts_2021Natur.589..207R,Trigg2024A&A...687A.173T}, and GRB\,231115A, associated with M82 \citep{Mereghetti2024Natur.629...58M,Trigg2025A&A...694A.323T}. 

    A pressing question is how frequently these events occur throughout the cosmos. This frequency can be expressed as a volumetric rate, the number of flares per unit volume per year. Estimating this rate is critical for understanding the population of MGFs, particularly given the small number of observed events. The first quantitative measurement of the volumetric rate of MGFs was presented by \citet{2021Burns}, who found a rate of $R_{\mathrm{MGF}} = 3.8^{+4.0}_{-3.1} \times 10^5~\mathrm{Gpc^{-3}~yr^{-1}}$ above a minimum isotropic-equivalent energy of $E_{\rm{iso,min}}=3.7\times10^{44} \rm erg$. Understanding the rate of MGFs offer direct insights into magnetar energy release mechanisms, as their temporal structure and energetics reflect processes such as magnetic reconnection and crustal failure induced by internal magnetic stresses \citep{ThompsonDuncan1995, ThompsonDuncan1996}. These signatures constrain models of how energy is transferred from the stellar interior to the magnetosphere and radiated away.

    The study of MGFs is also tightly linked to other astrophysical phenomena. Theoretical models predict that MGFs may emit gravitational wave (GW) signals, either as bursts during the flare onset or via excitation of NS normal modes \citep{abbott2007search,LIGOScientific:2019ccu, Macquet:2021eyn,Beniamini_2025}. These modes correspond to large-scale oscillations of the NS, including crustal shear vibrations and global seismic activity, which can be triggered by the sudden reconfiguration of the magnetic field or fracturing of the solid crust during an MGF. Such events involve rapid, asymmetric motions of dense matter, producing the time-varying quadrupole moment necessary for GW emission. Next-generation GW observatories \citep{Punturo2010} may detect such signals, offering unique insight into the NS interior and crust, with implications for constraining the equation of state of dense matter \citep{Patra_2020}. 
    
    MGFs have also been shown to be responsible for the formation of heavy elements like gold, platinum, and uranium via the rapid neutron capture process \citep[\textit{r}-process;][]{Cameron_osti_4709881,BBFH_RevModPhys.29.547}. The MGF from SGR\,1806-20 has been shown to produce conditions suitable for $r$-process nucleosynthesis: baryon-loaded ejecta expelled during a flare can undergo rapid neutron capture as it decompresses, particularly in the fast outer layers where a hot, $\alpha$-rich freeze-out enables the formation of both light and heavy nuclei \citep{Cehula_10.1093/mnras/stae358,patel2025direct}. Although unlikely to dominate Galactic heavy element production, such flares could contribute to early chemical enrichment given their short delay times after star formation. The early production of heavy elements may have played a critical role in sustaining molten planetary cores through radiogenic heating \citep{Stevenson_2008Natur.451..261S}, which are essential for the generation of magnetic fields. Such fields, in turn, help shield planetary atmospheres and create conditions favorable for the emergence of life.
    
    Magnetars are also associated with some fast radio bursts (FRBs), as demonstrated by FRB\,200428 from SGR\,1935+2154—a Galactic magnetar that emitted a coincident radio burst and X-ray outburst \citep{Popov_2018, bochenek2020fast}. However, this event was not an MGF, but rather a lower-energy outburst from a magnetar. No confirmed MGFs, Galactic or extragalactic, have yet shown associated FRB emission, even when simultaneous radio observations were available \citep{CHIME_FRB2023GCN.35070....1C}. Establishing whether MGFs can, or routinely do, produce FRBs remains a key question in understanding the origin of these millisecond-duration radio flashes.
    
    By building and characterizing a larger population of MGFs, we can place tighter constraints on their rates, energetics, and possible magnetar progenitors. Expanding the sample of MGFs reduces Poisson uncertainty in measuring their intrinsic rates and improves constraints on their energetics distribution. These improvements enhance the precision of forward-folding models, provide data to help resolve parameter correlations, and yield more robust insight into the physical processes underlying MGF production and magnetar evolution. By doing so, we can gain a better understanding of how magnetars form, how often they flare, and under what conditions they might produce related transients, such as FRBs. A more complete MGF sample will support the design of future GRB instruments and inform multi-messenger searches by contextualizing potential gravitational-wave or neutrino signals.

    We conduct a systematic archival search of the Fermi Gamma-ray Burst Monitor \citep[GBM;][]{meegan2009fermi} sGRB dataset, guided by an empirical evaluation of the known extragalactic MGF population (Section~\ref{sec:gbm_search}). This search was motivated by the prior identification of GRB\,180128A in an initial exploratory search \citep{Trigg2024A&A...687A.173T}, demonstrating that GBM could reveal additional low-fluence events not previously identified as MGFs. Then, building on the forward-folding population analysis of \cite{2021Burns}, we integrate these newly identified GBM candidates with previously published IPN events to construct a more complete sample. This combined dataset enables improved statistical constraints on the volumetric rate, minimum energetics, and energy distribution of MGFs while accounting for instrument-specific detection probabilities and spectral variation. By unifying GBM and IPN detections within a self-consistent Markov Chain Monte Carlo (MCMC) framework, our approach provides the most robust characterization of the extragalactic MGF population to date (Section~\ref{sec:pop_anal}). Section~\ref{sec:disc} discusses the implications of our results for magnetar formation, energetics, and recurrence, while Section~\ref{sec:conclude} summarizes the key conclusions and outlines future directions for the study of extragalactic MGFs.

\section{Archival Data Search for Extragalactic Magnetar Giant Flares}
\label{sec:gbm_search}

    In our refined search for extragalactic MGFs in archival data, we maintain our focus on the \fgbm. \fgbm is comprised of 12 uncollimated thallium-doped sodium iodide (NaI) detectors and two bismuth germanate (BGO) detectors, mounted on opposite sides of the spacecraft to ensure coverage of the full, unocculted sky at higher energies. Each NaI detector is sensitive to photons in the $\sim$8--900\,keV range, while the BGO detectors cover a higher-energy range of approximately 0.25--40\,MeV. Together, the NaI and BGO enable the detection and spectral characterization of a wide variety of gamma-ray transients.

    GBM continuously monitors the count rates in each detector and triggers when a statistically significant rate increase is observed in at least two NaI detectors. The on-board trigger threshold corresponds to a flux of roughly a few photons\,cm$^{-2}$\,s$^{-1}$ in the 50--300\,keV band (for a 1\,s peak interval). The sensitivity of \fgbm in the 50--300\,keV band is sufficient to detect the vast majority of GRBs, including weaker short events. We focus on the Time-Tagged Event (TTE) data, which records the time and energy of each photon detection during a burst with microsecond timing precision. Each TTE event includes an energy channel (out of 128 channels) and an absolute time stamp with 2\,$\mu$s resolution.
     
    The fine temporal and spectral resolution of \fgbm, combined with its broad sky coverage and high sensitivity to sGRBs, offers a unique opportunity to search for extragalactic MGFs hidden among the broader population of sGRBs. GBM is the most prolific detector of short GRBs, providing an unparalleled dataset for studying these events. Although \fgbm is not instantaneously all-sky, typically covering about two-thirds of the sky at any given moment, it continuously monitors a large fraction of the sky with high sensitivity. This enables the detection of bright short-duration events across a significant cosmological volume, making GBM an ideal instrument for archival MGF searches.

    However, given that MGFs account for only $\sim2\%$ of the total sGRB sample \citep{2021Burns}, directly attempting to associate every detected sGRB with nearby galaxies would be inefficient and would result in enormous losses of statistical power. To address this, we apply a down-selection procedure that filters for bursts with characteristics expected of extragalactic MGFs before attempting precise localization. This targeted approach reduces the population of interest by up to an order of magnitude, improving the sensitivity and efficiency of follow-up searches for host galaxy associations while mitigating false positives.

    Guided by characteristic observational signatures, we systematically scour the archival \fgbm dataset for previously unidentified MGFs. Here, we describe the data and methods used to discern MGF candidates from the broader population of GBM triggers.

\subsection{Refined Search}
\label{ref_search}

\begin{figure*}
        \centering
        \includegraphics[width=0.99\linewidth]{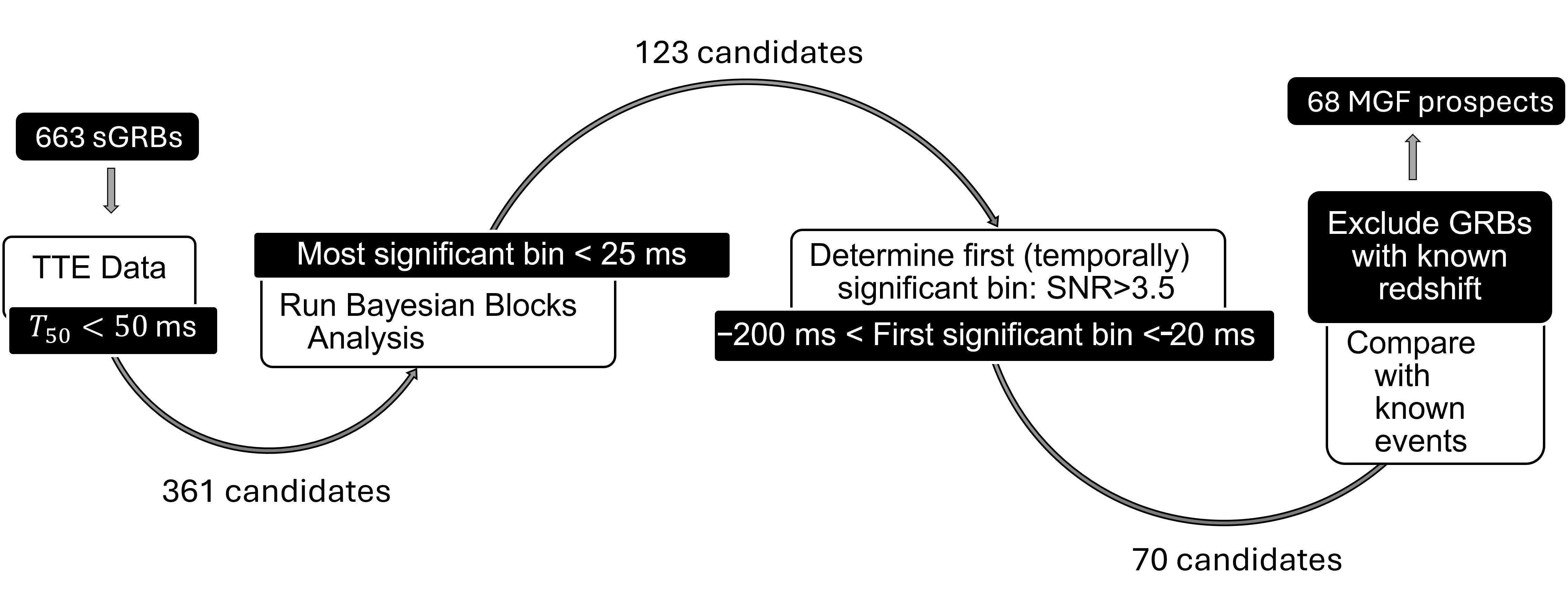}
        \caption[Flowchart illustrating the refined selection criteria used to identify MGF candidates from the \fgbm sGRB sample.]{Flowchart illustrating the refined selection criteria used to identify MGF candidates from the \fgbm sGRB sample. Starting from 663 sGRBs with available TTE data, a series of temporal cuts is applied using BB analysis. GRBs with known redshift are excluded, yielding 68 MGF prospects.}
        \label{fig:refined}
    \end{figure*}

    With GRB\,180128A and GRB\,231115A now added to the sample of known extragalactic MGFs, we turn to refining selection criteria to search for similar events within the \fgbm catalog more robustly. This refinement required accounting for both the observed properties of the MGF population and the temporal and spectral limitations of the observing instruments. We utilize a Bayesian Blocks-based approach \citep[BB;][]{Scargle2013bayesian} as a temporal analysis selection filter, focusing on rise time and duration. For our revised search, we analyze all GBM triggers that occurred prior to March 2025 using TTE data available in the \fgbm catalog. To clarify the event selection process, Figure~\ref{fig:refined} presents a flowchart detailing the selection criteria and the number of bursts that pass each step.

    To calibrate our selection criteria for extragalactic MGFs, we require a representative sample of confirmed events. Given that \fgbm has only observed three extragalactic MGFs (GRB\,200415A, GRB\,180128A, and GRB\,231115A) we must look to another instrument that has high-quality observations of the remaining three bursts. The best such instrument is \kwind \citep{KONUS1995SSRv...71..265A}, which has detected and characterized multiple confirmed extragalactic MGFs. Since its launch in 1994, \kwind has provided all-sky gamma-ray coverage using two NaI detectors oriented toward the ecliptic poles. In triggered mode, the instrument records high-resolution light curves in three energy bands, G1 (18–70\,keV), G2 (70–300\,keV), and G3 (300–1100\,keV), over a $\sim$230\,s window spanning from $T_0 - 0.512$\,s to $T_0 + 229.632$\,s. The time resolution steps from 2\,ms to 256\,ms over the course of the burst, and 64 multichannel spectra are acquired from 10\,keV to 10\,MeV with adaptive integration times \citep{KONUS1995SSRv...71..265A, svinkin2021bright}.

    We define the Konus sample as the four confirmed extragalactic MGFs observed by \kwind for which we have data: GRB\,051103, GRB\,070201, GRB\,070222, and GRB\,200415A \citep{svinkin2016ApJS..224...10S,svinkin2021bright}. For GRB\,200415A, which GBM also observed, we utilize the \kwind data to maintain consistency across the sample. The inclusion of high-quality \kwind data allows us to refine our selection filters based on well-characterized MGF properties. To quantify the temporal structure of each event, we applied a BB algorithm to the \kwind light curves. This produced consistent, instrument-independent characterizations of each burst’s timing behavior.

    Given that \( N_{\mathrm{photons}} \propto 1/d^{2} \) when fixing intrinsic brightness, we expect the number of detected counts for a given sGRB to decrease with distance. As the counts diminish, the BB analysis becomes more likely to identify fewer significant bins, often resulting in longer-duration bins where signal significance is reduced. To empirically determine appropriate temporal cuts, we took the full sample of known extragalactic MGFs (from both the GBM and Konus datasets). We attenuated their photon fluxes from the distances to their respective host galaxies to a distance of 50\,Mpc, in 0.5\,Mpc increments. Our burst attenuation analysis allowed us to evaluate the distance at which our BB-based detection algorithm would fail to identify the known MGFs. We defined failure as the presence of fewer than three BB bins (i.e., when no statistically significant temporal structure remains detectable). However, our analysis also shows that as distances increase by an order of magnitude, the BB binning for MGFs can appear as bright, isolated spikes, with the most significant bin typically spanning 10–30\,ms. 
    
    Beginning our search of GBM archival data, we use the {\tt bcat detector mask}\footnote{\url{https://heasarc.gsfc.nasa.gov/w3browse/fermi/fermigbrst.html}}, which reports the good detectors for each burst as determined by the GBM team. Good detectors are defined as NaI detectors with a boresight angle within 60 degrees of the source. The {\tt bcat} files provide basic burst information such as duration, peak flux, and fluence, and include only data from these good-viewing detectors. We then select the appropriate BGO detector based on which side of \Fermi has the greater number of selected NaI detectors. We adopt a uniform bin size of 2~ms for all light curves. This resolution preserves the prompt peak while allowing us to probe finer temporal structure within the burst.

    The first parameter we examined was the $T_{50}$ duration. The $T_{50}$ is the duration over which 50\% of the burst fluence accumulates, starting when 25\% of the total fluence is detected, measured between 50 keV and 300 keV. However, when applying this cut to the full GBM trigger catalog of sGRBs, we found that significantly fewer bursts from earlier in the mission satisfied this criterion. This discrepancy arises from evolving choices in how finely burst durations were evaluated--particularly the temporal resolution of the light curves used to compute $T_{50,GBM}$. In the early catalog, the fixed time binning sometimes exceeded the actual burst duration, which limited the precision of duration measurements for the shortest bursts. To ensure consistent and accurate treatment of $T_{50}$ durations across the entire dataset, we recalculated the durations ($T_{50,BB}$) for the full sGRB sample using a BB–based approach, similar to the {\tt battblocks}\footnote{\url{https://heasarc.gsfc.nasa.gov/docs/software/lheasoft/help/battblocks.html}} algorithm used to estimate burst durations for the \swift Burst Alert Telescope \citep{Barthelmy05}.
    
    We start by selecting events in the full catalog with $T_{50, GBM} \leq 250\,\mathrm{ms}$. This duration was chosen due to no known MGF having a full duration beyond $\sim$150\,ms, allowing us to reduce the initial sample considered from 663 sGRBs to 361 bursts. To improve accuracy, we apply partial-bin interpolation when determining \(T_{50, BB}\) durations. This method assumes a constant count rate within each BB bin. If the cumulative counts cross the 25\% or 75\% thresholds that define the $T_{50}$ within a bin, we solve for the precise time at which that threshold is reached, rather than assigning all counts to the bin edge. At the millisecond timescales relevant to our analysis, this approach yields a more robust estimate of \(T_{50}\) than simple linear interpolation. Examining the distribution of \(T_{50, BB}\) durations among known MGFs, we observe that none exceed $\sim40\,\rm{ms}$. Thus, we conservatively impose a selection criterion of \(T_{50, BB}\leq50\,\rm{ms}\).

    The next parameter we set out to refine was the duration of the bright, main spike of the burst. This corresponds to the BB bin with the highest count rate per unit time, referred to as the most significant BB bin. To account for the results of our attenuation analysis, we first include in our sample any burst that displays a single significant BB bin, as such cases may represent distant MGFs for which reduced photon statistics obscure finer structure. For all other bursts, we track the peak duration (as measured by the most significant BB bin) as a function of distance. The attenuation analysis also revealed that, out to 9.5\,Mpc, all events in the sample maintained a peak duration of $\leq25\,\rm{ms}$. Beyond this distance, only GRB\,231115A exceeded this threshold, with a peak duration of 42\,ms at 20\,Mpc; a distance within the detectable range for \fgbm, but beyond the range where events can typically be confidently identified due to localization limitations. Based on these results, we adopted a peak duration cut of 25~ms for multi-BB-bin bursts. These considerations result in a sample of 123 prospective candidates.
        
    To ensure we only select events with sharp rise times similar to those seen in the extragalactic MGF population the final parameter under consideration is the first significant bin (FSB). The initial criterion--removing bursts with a bin of signal-to-noise ratio~$\geq$3.5 occurring more than 10\,ms before the main peak--was intended to select events with fast rise times, consistent with empirical observations of known MGFs, where the most significant pulse typically occurs first. We therefore retained the FSB-based cut to exclude bursts with early-rising components inconsistent with the observational and theoretical profile.
    
    We therefore implement a selection criterion that excludes any burst containing a significant bin between 200\,ms and 20\,ms prior to the main peak. This approach allows us to preserve potential MGFs exhibiting precursor emission while ensuring that only bursts with sharp rise-times are selected. Applying this criterion to our down-selected sample, we are left with 70 remaining prospective MGF candidates. We then manually excluded two bursts: GRB\,150101B, which is a likely NS-NS merger \citep{Burns_2018}, and GRB\,150101A, which exhibited an X-ray afterglow and is associated with a host galaxy at $z=0.426$, confirming its cosmological origin. This refinement resulted in a final sample of 68 MGF candidates.

\subsection{Galaxy association}

    To assess the likelihood that each of the 68 MGF candidates originated from a nearby star-forming galaxy, we implemented the False Alarm Rate (FAR) significance test described in \cite{2021Burns} \citep[see also][]{Negro2023IAUS..363..284N, Trigg2024A&A...687A.173T}. For each GRB, we computed a ranking statistic, $\Omega$, defined as the probability-weighted overlap between the GRB localization probability distribution $P_i^{\mathrm{GRB}}$ and the expected distribution of extragalactic MGFs, $P_i^{\mathrm{MGF}}$. That is, $P_{i}^{MGF} \propto {\rm SFR} \times \rm{PDF(\textit{E}_{iso}}(\textit{D,S}))$ where $D$ is the host galaxy distance and $S$ the fluence. Each local galaxy, taken from \citet{z0mgs_2019}, is weighted linearly by its star-formation rate (SFR), equally distributed over its apparent size, and by the prevalence of MGFs, which is inferred from the $E_{iso}$ from the given galaxy distance and GRB fluence pairing. All galaxies in the sample are placed on the sky to create $P_{i}^{MGF}$. The resulting $\Omega$ values were then compared to a background distribution created from $45,000$ randomized galaxy distributions, preserving local structure and global SFR distributions.

\begin{figure*}[ht!]
        \centering
        \includegraphics[width=0.95\textwidth]{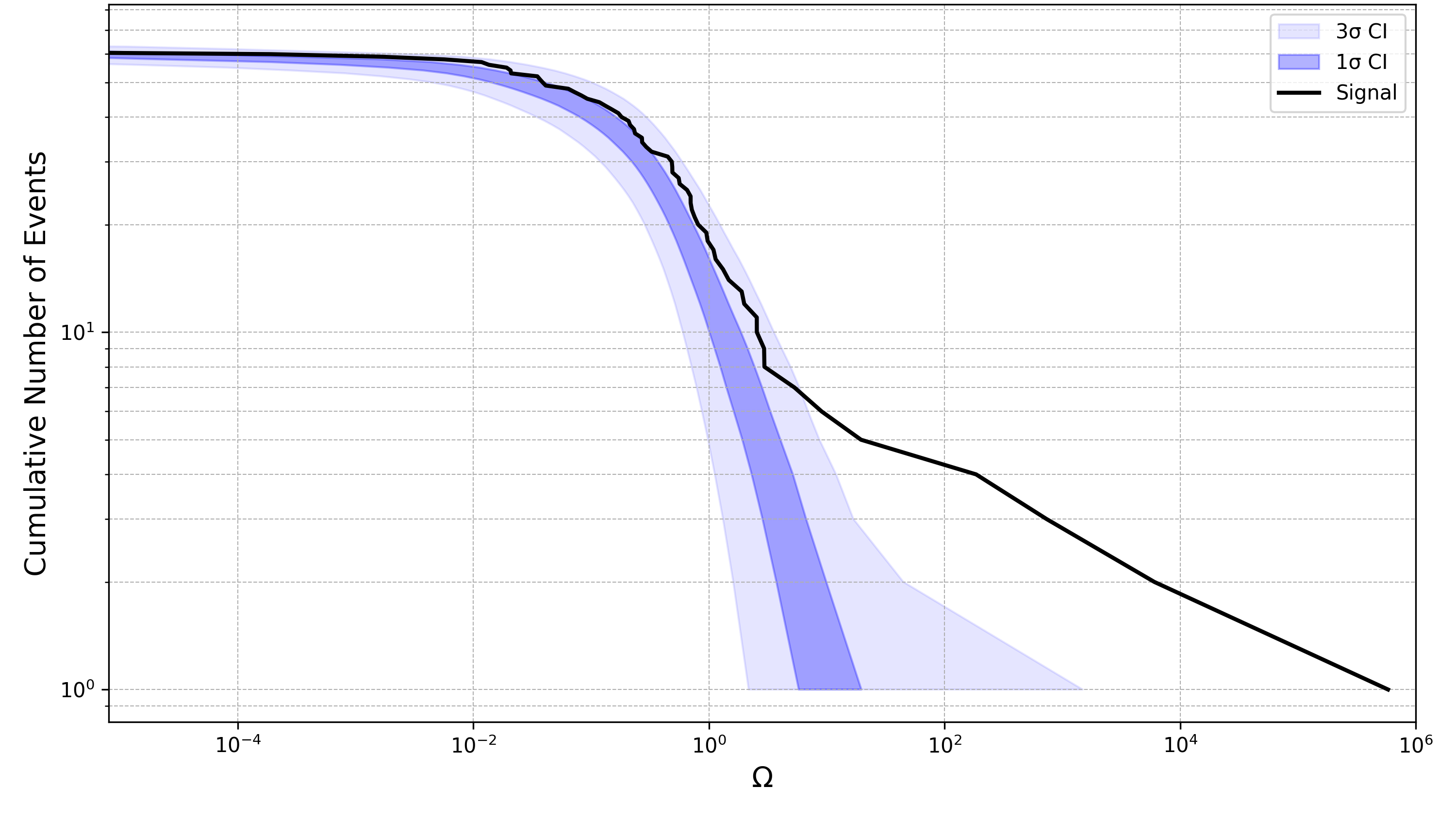}
        \caption[Cumulative $\Omega$ values for the 68 MGF candidates (black line) compared to the background distribution generated from random galaxy rotations.]{Cumulative $\Omega$ values for the 68 MGF candidates (black line) compared to the background distribution generated from random galaxy rotations. The shaded regions indicate the $1\sigma$ (dark blue) and $3\sigma$ (light blue) confidence intervals derived from resampling 10,000 background comparisons drawn from an initial set of 45,000 randomized galaxy realizations.}
        \label{fig:gbm_significance}
    \end{figure*}

    As shown in Figure~\ref{fig:gbm_significance}, which shows the distribution of candidate $\Omega$ values (black curve), we define candidate significance based on their location relative to the 99.73\% background exclusion contour, corresponding to a 3$\sigma$ (light blue) and 1$\sigma$ (dark blue) contour separation line. These confidence intervals are derived from the randomized background trials, following the methodology for generating random realizations that preserve spatial statistical properties as outlined in \citet{2021Burns}. Using the cleanest separation region in $\Omega$ space, we identify a cut near the boundary, providing a statistically motivated threshold for classifying events as strong MGF candidates. After assigning host galaxies to the full candidate sample using the method described above, we repeat the host association procedure for each of the bursts above the 3$\sigma$ threshold individually. This step ensures that associations for the most significant events are robust and not biased by bulk-sample assumptions.
    
    A summary of the top-ranked bursts, their host association significance percentages, $\Omega$ values, and associated spectral parameters is provided in Table~\ref{tab:mgf_candidates}. The spectral fits were performed using a time-integrated Comptonized \citep[COMPT;][]{gruber2014fermi} model, or drawn from published values where available. The COMPT function exhibits a power-law behavior characterized by an index $\alpha$, with an exponential cutoff at a characteristic energy $E_{\rm{p}}$ of the spectral peak of a $\nu F_{\nu}$ representation.

    GRB\,231115A, GRB\,200415A, and GRB\,180128A, three known extragalactic MGFs, are robustly recovered in this analysis, each with $\Omega$ values well above the 3$\sigma$ false alarm threshold. Their temporal and spectral properties remain consistent with expectations for MGF emission and provide an important benchmark for candidate evaluation.

    An additional four bursts exhibit $\Omega$ values that fall just outside the 3$\sigma$ confidence region of the population distribution, indicating that while their galaxy associations are less significant than the strongest candidates, these track star-forming regions more than expected by chance. Each of these bursts is well fit by a COMPT spectral model (Table~\ref{tab:mgf_mcmc_sample}), consistent with the behavior of confirmed extragalactic MGFs. The values for the spectral models, three of which are taken from existing catalogs and one obtained from a spectral analysis performed in this work (GRB\,200423A), all exhibit $E_{\rm{p}}$ and $\alpha$ values within the range observed for the known population, further supporting their classification as plausible MGF candidates. 
    
\renewcommand{\arraystretch}{1.2}
\begin{table*}[ht!]
\setlength{\tabcolsep}{3.pt}
    \caption{Summary of MGF Candidates and Host Galaxy Associations}
    \label{tab:mgf_candidates}
    \centering
    \begin{tabular}{lcccccc}
        \hline\hline
        GRB Name & $\Omega$ & Associated & Distance & Association & Fluence & $E_{\mathrm{iso}}$ \\
         & & Galaxy & (Mpc) & (\%) & ($\times10^{-7}~\rm{erg\,cm}^{-2})$ & ($\times10^{44}~\rm{erg}$) \\
        \hline
        GRB\,231115A  & 581,513 & M82 & 3.5  & >99.9  & $7.8\pm0.3$ & $11.5\pm0.4$ \\
        GRB\,200415A  & 6041 & NGC\,253    & 3.7  & >99.9  & $97\pm3$ & $142\pm5$ \\
        GRB\,180128A  & 734 & NGC\,253    & 3.7  & >99.9   & $4.1\pm0.9$ & $6\pm1$ \\
        GRB\,120616A  & 185 & IC\,0342    & 2.3  & 98   & $3.6\pm0.3$ & $2.3\pm0.2$ \\
        GRB\,200423A  & 20 & NGC\,6946   & 7.7  & >99.9     & $12\pm3$ & $85\pm20$ \\
        GRB\,231024A  & 9 & NGC\,253    & 3.7  & 94     & $3.4\pm0.4$ & $5.5\pm0.7$ \\
        GRB\,081213  & 5 & NGC\,253    & 3.7  & 96     & $2.6\pm0.2$ & $4.2\pm0.3$ \\
        \hline
    \end{tabular}
    \tablefoot{GRB\,231115A, GRB\,200415A, and GRB\,180128A are the three known extragalactic MGFs detected by \fgbm. Association percentages represent the fraction of the total $\Omega$ statistic contributed by the listed host galaxy. Fluence values are in the bolometric range (e.g., 1~keV--10~MeV)}
\end{table*}
        
    Taken together, these results reinforce the growing body of evidence that a small but significant fraction of sGRBs detected by \fgbm originate from extragalactic MGFs. Given that our sample consists of numerous bursts with extremely poor localizations, which preclude them from consideration as MGF candidates, we estimate a lower limit of 1.1\%. This result is consistent with the previously reported 2\% estimate by \cite{2021Burns}. The strongest candidates exhibit both high $\Omega$ values and spectral characteristics consistent with MGF emission.

\begin{table*}[ht!]
    \caption{Summary of Bursts Used in MCMC Modeling}
    \label{tab:mgf_mcmc_sample}
    \centering
    \begin{tabular}{lcccc}
        \hline\hline
        GRB Name & $E_{\mathrm{iso}}$ ($10^{44}$ erg) & $E_{\mathrm{p}}$ (keV) & $\alpha$ & Sample\\
        \hline
        GRB\,790305B & $\sim$1.2 & -- & -- & NA \\
        GRB\,980827 & 4.3 & -- & -- & Burns/IPN \\
        GRB\,041227 & 230 & 850 & -0.7 & Burns/IPN \\
        GRB\,051103 & 530 & 2150 & -0.2 & Burns/IPN \\
        GRB\,070201 & 16 & 280 & -0.6 & Burns/IPN \\
        GRB\,070222 & 62 & 1290 & -1.0 & Burns/IPN \\
        GRB\,200415A & 142 & 1080 & -0.0 & Burns/IPN/GBM\\
        GRB\,180128A & 6.0 & 290 & -0.6 & IPN/GBM \\
        GRB\,231115A & 11.5 & 600 & -0.1 & IPN/GBM \\
        GRB\,120616A & 2.3 & 500 & -0.4 & GBM \\
        GRB\,200423A & 85 & 600 & -0.7 & GBM \\
        GRB\,231024A & 5.5 & 500 & -0.8 & GBM \\
        GRB\,081213 & 4.2 & 400 & -0.6 & GBM \\
        \hline
    \end{tabular}
    \tablefoot{Values for $E_{\mathrm{iso}}$ are computed assuming isotropic emission and host galaxy distances. Spectral parameters ($E_{\mathrm{p}}$, $\alpha$) are drawn from time-integrated COMPT model fits where available. The Burns sample refers to all events analyzed in \citet{2021Burns}. The IPN sample extends this set by including two additional bursts identified subsequently. The GBM sample comprises all events detected by \fgbm, including the four newly reported here.}
\end{table*}
    
    Our initial expectation was that an archival search would primarily recover fainter MGF candidates--those missed during real-time analysis due to marginal fluence or poor localization. This was broadly borne out: many of the newly identified events exhibit lower fluence and simplified temporal structure, consistent with distant or sub-threshold flares. However, a subset of the bursts are sufficiently bright that they may warrant follow-up searches for additional detections for possible localization improvement.

    While none of the four newly identified events can be unambiguously classified as MGFs, they are consistent with expectations for extragalactic flares and significant on a population level, expanding the available sample for population inference. These findings provide a statistically robust foundation for constraining the volumetric rate, energetics, and PL distribution of the extragalactic MGF population, as explored in Section~\ref{sec:pop_anal}. The complete set of MGFs used in this analysis is outlined in Table~\ref{tab:mgf_mcmc_sample}, which forms the basis for all subsequent modeling.

\section{Population Analysis}
\label{sec:pop_anal}

    Constraining the volumetric rate and minimum isotropic-equivalent total energy release for extragalactic MGFs requires a forward-folding modeling framework that accounts for instrumental sensitivity, spectral diversity, and selection effects. Observed samples of MGF candidates are sparse and heterogeneous, drawn from instruments with differing sky coverage, sensitivities, and triggering algorithms. These limitations make population-level inference challenging, particularly when attempting to estimate intrinsic properties from detections subject to severe and energy-dependent selection biases. Moreover, while the intrinsic energetics distribution of MGFs is most naturally expressed in terms of isotropic-equivalent total energy, detector thresholds and recovery efficiencies are fundamentally tied to photon space, as these are photon-counting instruments. The significant spectral variation observed across different MGFs must be taken into consideration when translating between intrinsic properties and observables.

    The initial effort by \citet{2021Burns} introduced such a forward-folding modeling approach. That study simulated synthetic MGF populations originating from galaxies within 5\,Mpc, using a galaxy catalog to draw source locations, with $E_{\rm iso}$ drawn from a PL distribution and event rates scaled to each galaxy’s SFR. Detectability was evaluated using a fixed energy fluence threshold representative of generalized IPN sensitivity, and sample completeness was assessed accordingly. While this approach marked a critical step forward, it avoided consideration of spectral variation entirely by remaining in energy fluence space and not converting to photon space. As a result, all bursts were implicitly treated as spectrally identical in terms of detectability. It also approximated instrumental sensitivity using a single energy fluence cutoff, and separated the analysis of rate, spectral shape, and detectability into distinct steps, precluding fully self-consistent parameter inference.
    
    To address these limitations, we developed an MCMC implementation that performs posterior inference on key population parameters, the rate of MGFs per solar mass of star formation--or SFR-scaled rate--$R$, minimum total energetics $E_{\rm{iso, min}}$, and PL index $\beta$ for the various samples. For the GBM sample we incorporate a more realistic and physically motivated treatment of spectral variation and detection probability.

\subsection{MCMC Modeling: IPN Implementation and Comparison to Burns}
\label{sec:MCMC_model}

    This section describes the MCMC model in the context of reproducing the results of \citet{2021Burns} using the same IPN sample and assumptions.

    The model begins by assuming that the distribution of $E_{\rm iso}$ for MGFs follows a PL ($\beta$) form, where the probability of drawing a given $E_{\mathrm{iso}}$ scales as $E_{\mathrm{iso}}^{-\beta}$ between a lower cutoff $E_{\rm iso, min}$ and a fixed maximum energy $E_{\rm iso, max}=5.75\times10^{47}$\,erg \citep{2021Burns}. The lower cutoff is treated as a free parameter in the model and represents a hard threshold below which there are no events, rather than a soft or exponential turnover. At each MCMC iteration, new values for the $\beta$, $E_{\rm iso, min}$, and $R$ (expressed in MGFs per $M_\odot$) are sampled from their respective priors. Using these parameters, a synthetic population of MGFs is generated by first drawing $E_{\mathrm{iso}}$ values from the PL distribution. Each burst is then assigned a host galaxy selected with probability proportional to its SFR, using the z=0 Multiwavelength Galaxy Synthesis Catalog \citep{z0mgs_2019} limited to a set maximum distance. Finally, the bolometric energy fluence is calculated by assuming isotropic emission and applying inverse-square law scaling, such that $S=E_{\mathrm{iso}}/(4\pi D^2)$, where $D$ is the distance to the host galaxy.

    Detection is defined as a sharp energy fluence threshold of $2\times10^{-6}\,\rm{erg\,cm^{-2}}$, consistent with the approach of \citet{2021Burns}. Simulated events with bolometric fluence (1\,keV--10\,MeV) above this threshold are marked as detected, while those below are discarded, and IPN is treated as effectively all-sky. The expected number of detections is computed based on the observation time ($T_{\rm{obs}}$), completeness, and SFR distribution.

    To perform the parameter inference, we use the \texttt{emcee} Python package \citep{ForemanMackey2013}, an affine-invariant ensemble sampler that efficiently explores correlated and high-dimensional parameter spaces. We configure the sampler with 80 walkers and 15,000 steps per walker, discarding the first 1,500 steps as burn-in. Walkers are initialized near literature values, with priors drawn from narrow Gaussian distributions centered at $\log_{10} (R*M_{\odot}) \sim -1.3\,$, $\log_{10}( E_{\rm{iso,min}}) \sim 43.8\,\rm{erg}$, and $\beta = 1.7$. Convergence is assessed using the acceptance fraction, autocorrelation time, and visual inspection of the posterior distributions via \texttt{corner}\footnote{\url{https://corner.readthedocs.io/en/latest/}}\,\citep{corner} plots.

    As in \cite{2021Burns}, we cannot directly infer a volumetric rate due to the known overdensity of galaxies in the local universe. To account for this and enable comparison with a homogeneous SFR volume, we use the total measured star formation rate within 50\,Mpc to convert $R$ into a volumetric rate above a given $E_{\rm{iso,min}}$. 
    
    We apply updated priors (described below) and observation time to the same IPN sample analyzed in \citet{2021Burns}. Adopting an observation start date of 21 April 1991, the beginning of concurrent BATSE and \textit{Ulysses} operations, which together enabled extragalactic burst localization, we take $T_{\rm{obs}} = 29$,yr. The previously reported values and those obtained with our MCMC method are listed in Table~\ref{tab:mcmc_results}.

    Priors on the parameters are selected to reflect both physical intuition and empirical constraints, and are implemented as probability distributions within broader hard bounds enforced by the sampler. For the SFR-scaled rate $R$, we adopt a uniform prior in log-space over the range $\log_{10} (R*M_{\odot}) = [-3, 6]$. The lower limit is motivated by the estimated MGF rate in the Milky Way, while the upper bound exceeds the highest previously inferred extragalactic rates by several ordered of magnitude. For the minimum energy $E_{\rm{iso,min}}$, we use a uniform prior spanning $10^{42.5}$\,erg to $10^{44.1}$\,erg. This range includes typical energies of lower-energy recurrent magnetar emissions on the low end, and is truncated at the lowest $E_{\mathrm{iso}}$ measured in the observed sample \citep[GRB\,790305B;][]{mazets1979observations} on the high end. For the PL index $\beta$, we adopt a Gaussian prior centered at 1.7 with a standard deviation of 0.4, with hard bounds at 1.3 and 2.5. This range reflects values explored in \citet{2021Burns} and permits thorough exploration of plausible parameter space without imposing strong biases.
  
    In the five years since the publication of \citet{2021Burns}, additional extragalactic MGFs have been identified by the IPN, extending the baseline. We apply the model to this Updated IPN sample, which includes GRB\,180128A and GRB\,231115A and reflects increased temporal coverage. Incorporating these newer events provides a more complete view of the population and improves statistical power. While this expanded dataset spans multiple instruments and lacks a unified detection efficiency curve, we adopt the same fixed energy fluence threshold methodology as before to ensure consistency with the original analysis. We utilize $T_{\rm{obs}} = 34$\,yr for this Updated IPN sample. For the SFR distribution, we follow \citet{2021Burns} in including only galaxies within 5\,Mpc, as all events identified occurred within that volume.

\subsection{Modeling Improvements for GBM Sample}
\label{sec:improvements}    

    While the IPN-based implementation successfully reproduces prior results, it remains limited by simplifying assumptions regarding detection and spectral uniformity. To improve on this, we extend the MCMC framework to model the sample of MGF candidates detected by \fgbm. As described in Section~\ref{sec:gbm_search}, GBM’s greater sensitivity allows it to detect lower-energy bursts, enabling several key refinements to the population modeling. The GBM events used in this analysis represent previously identified MGFs and the new candidate identifications from this work. 
    
    First, we incorporate spectral variation across the population by assigning spectral parameters to each simulated burst. Specifically, the peak energy $E_{\mathrm{p}}$ is computed from the burst’s isotropic-equivalent energy $E_{\mathrm{iso}}$ using a best-fit relation derived from the observed $E_{\rm{p}}$--$E_{iso}$ distribution of the full MGF candidate sample (Figure~\ref{fig:Ep_Eiso_fit}), using the fitting method described in Appendix A of \citep{Trigg2025A&A...694A.323T}. The photon index $\alpha$ is assigned as a function of $E_{\mathrm{p}}$ using a linear mapping based on observed COMPT fits, wherein higher $E_{\mathrm{p}}$ values correspond to $\alpha \sim 0.0$ and lower $E_{\mathrm{p}}$ values to $\alpha \sim-1.0$. This theoretically motivated relationship captures the trend seen in known events and is implemented as a piecewise function that increases monotonically with $E_{\mathrm{p}}$ and flattens at high energies, ensuring spectral shapes that match observations across the simulated population.

\begin{figure}[ht!]
\centering
\includegraphics[width=0.9\linewidth]{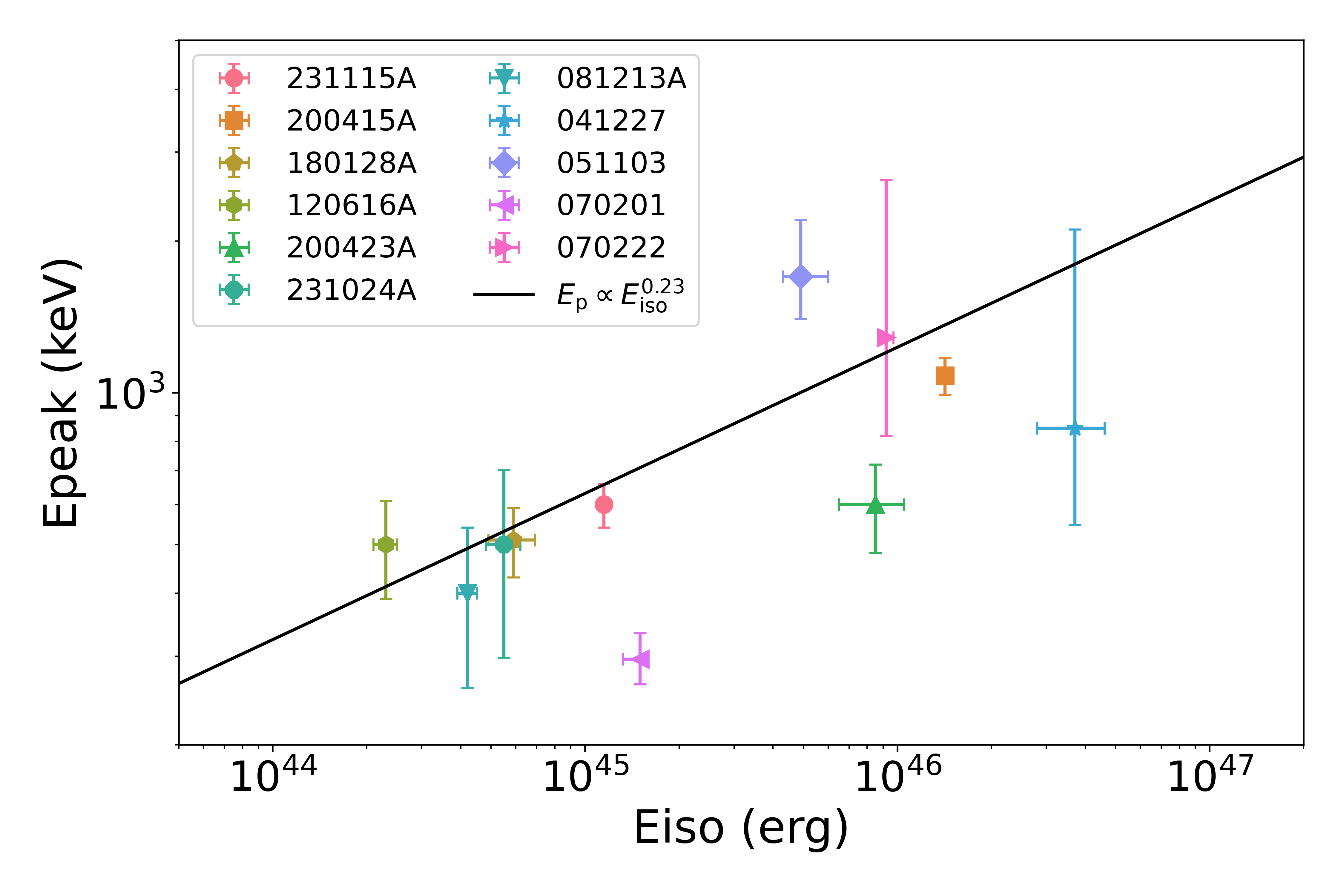}
\caption[Peak energy ($E_{\mathrm{p}}$) versus isotropic energy ($E_{\mathrm{iso}}$) for the sample of extragalactic MGF candidates and confirmed events.]{Peak energy ($E_{\mathrm{p}}$) versus isotropic energy ($E_{\mathrm{iso}}$) for the sample of extragalactic MGF candidates and confirmed events. Each burst is labeled and color-coded. Error bars reflect uncertainties in both $E_{\mathrm{iso}}$ and $E_{\mathrm{p}}$. The black line shows the best-fit empirical relation. This relation is used to assign $E_{\mathrm{p}}$ values in the population model as a function of $E_{\mathrm{iso}}$.}
\label{fig:Ep_Eiso_fit}
\end{figure}

\begin{figure}[ht!]
\centering
\includegraphics[width=0.9\linewidth]{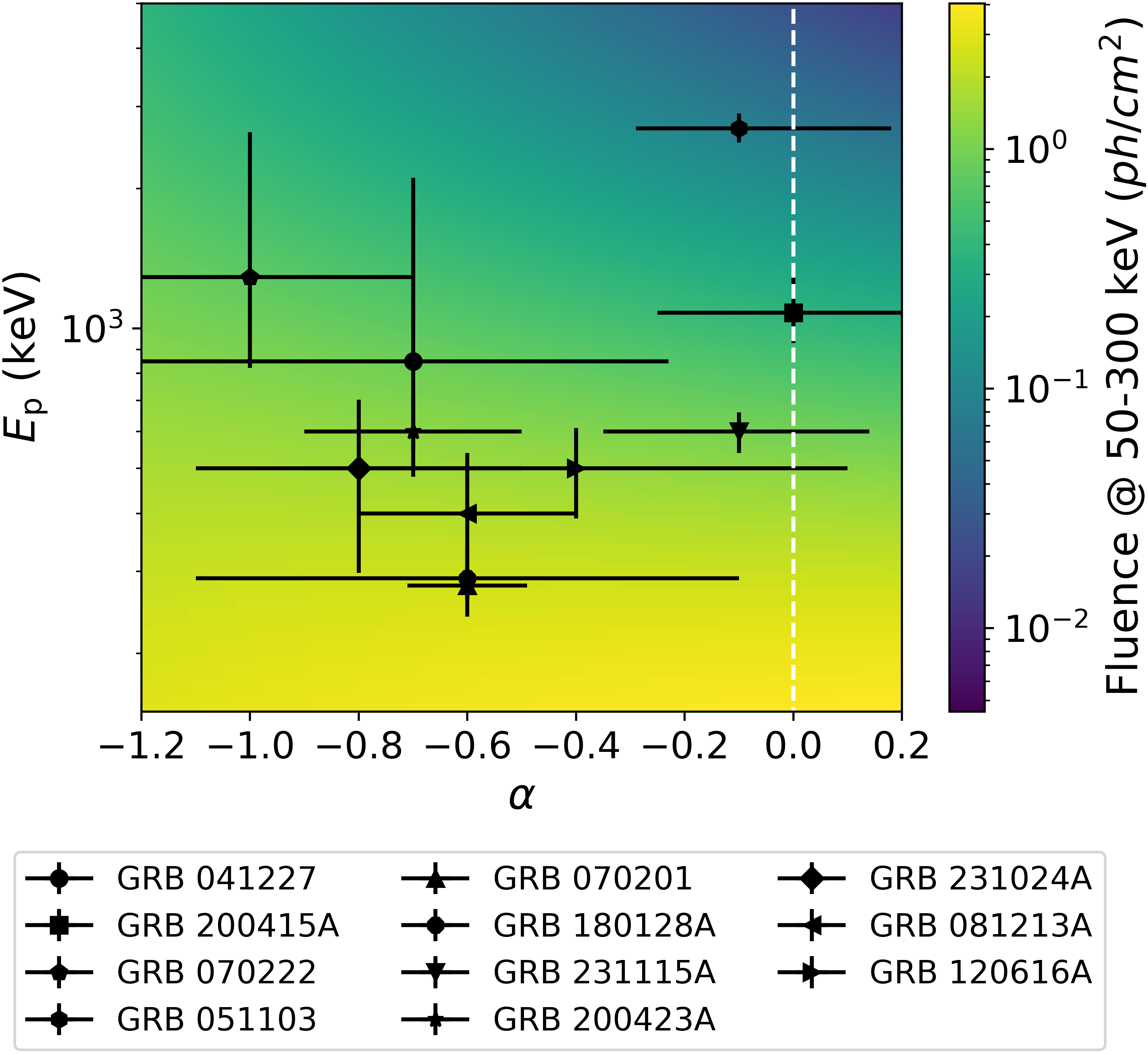}
\caption[Normalized fluence scaling surface at 50–300\,keV as a function of COMPT spectral parameters $\alpha$ and $E_{\mathrm{p}}$, constructed using a grid of model spectra and normalized to the observed fluence of GRB\,200415A.]{Normalized fluence scaling surface at 50–300\,keV as a function of COMPT spectral parameters $\alpha$ and $E_{\mathrm{p}}$, constructed using a grid of model spectra and normalized to the observed fluence of GRB\,200415A. Overlaid points correspond to observed MGF candidates. Units are relative and do not represent absolute fluence values.}
\label{fig:Ep_alpha_grid}
\end{figure}

    These spectral parameters are then used to compute photon fluence via a precomputed grid of photon fluence values, derived from COMPT models spanning the relevant $E_{\mathrm{p}}$ and $\alpha$ parameter space, as shown in Figure~\ref{fig:Ep_alpha_grid}. This grid, built from observed MGF spectra, is interpolated using \texttt{RegularGridInterpolator} from the Python package {\tt scipy}\footnote{\url{https://docs.scipy.org/doc/scipy/reference/generated/scipy.interpolate}} during MCMC sampling to efficiently convert $E_{\mathrm{iso}}$ to photon fluence as a function of both spectral shape and source distance. Photon fluences are then converted to peak photon fluxes assuming a 64\,ms integration window, consistent with GBM triggering timescale of GBM.
    
\begin{figure}[ht!]
    \centering
    \resizebox{\hsize}{!}{\includegraphics{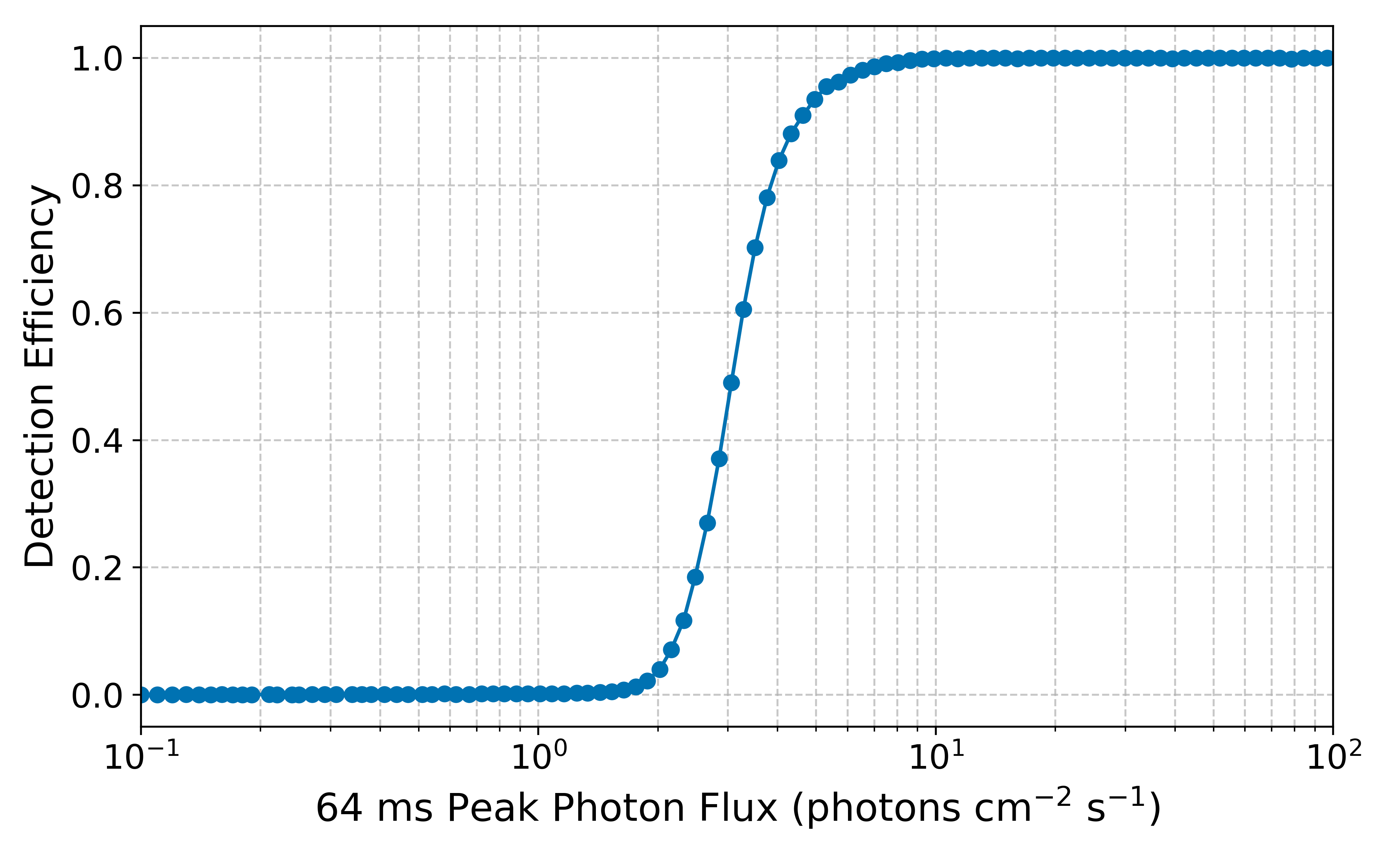}}
    \caption{GBM detection probability as a function of 64\,ms peak photon flux in the 50--300\,keV band.}
    \label{fig:GBM_sensitivity}
    \end{figure}

    Detection probabilities are derived from the GBM sensitivity curve (Figure~\ref{fig:GBM_sensitivity}) corresponding to its 64\,ms trigger window, under the assumption that the full $E_{\rm iso}$ is emitted within this timescale. While GBM employs multiple trigger windows, our efficiency model specifically reflects the 64\,ms response. This curve is interpolated to determine characteristic recovery thresholds (e.g., a 50\% detection threshold at approximately 3~ph\,cm$^{-2}$\,s$^{-1}$) over the energy range from 50\,keV to 300\,keV. Each event’s photon flux is compared with this sensitivity curve to determine the probability of detection, and stochastic realizations are drawn via uniform random sampling. Detection efficiencies are further scaled by GBM’s effective sky coverage and observing duty cycle, accounting for a 60\% exposure fraction due to limited field of view and periodic passage through the South Atlantic Anomaly, a region where inner radiation belt of Earth dips closest to the surface, resulting in elevated fluxes of energetic particles that expose satellites, including the ISS, to increased ionizing radiation. Additionally, a scaling based on the total observing time of $T_{\rm{obs}} = 16.72$ years is applied. The farthest event in the GBM sample, GRB\,200423A, is located at 7.7\,Mpc. Based on this, we conservatively adopt a SFR volume limited to 10\,Mpc for the GBM sample. While GBM may in principle be sensitive to extragalactic MGFs out to $\sim$25\,Mpc, this more restrictive threshold ensures consistency with the observed sample.

\subsection{Results from Individual Samples}

    To quantify the impact of sample selection on the inferred MGF population parameters, we ran independent MCMC analyses on each of the three main data sets: the \cite{2021Burns} IPN sample, the Updated IPN sample (which includes GRB\,180128A and GRB\,231115A), and the GBM candidate sample identified in this work. Each sample differs in completeness, spectral quality, and instrumental coverage, providing complementary windows into the underlying population.

\renewcommand{\arraystretch}{1.2}
\begin{table*}[ht!]
\centering
\caption{Posterior Parameter Estimates from MCMC Fits to Individual Samples}
\label{tab:mcmc_results}
\begin{tabular}{lccc}
\hline\hline
Sample & $R$ & $\beta$ & $E_{\rm{iso,min}}$ \\
      & ($M_{\rm{\odot}}^{-1}$)&  & ($\times10^{43}$\,erg) \\
\hline
Burns (2021) Results & $0.05^{+0.05}_{-0.04}$ & $1.7^{+0.4}_{-0.4}$ & 37 \\
Burns (2021) MCMC Benchmark & $0.05^{+0.14}_{-0.04}$ & $1.78^{+0.43}_{-0.37}$ & $16^{+24}_{-15}$ \\
Burns (2021) IPN MCMC & $0.07^{+0.16}_{-0.05}$ & $1.64^{+0.34}_{-0.30}$ & $4.2^{+6.8}_{-3.5}$ \\
Updated IPN MCMC & $0.09^{+0.16}_{-0.06}$ & $1.67^{+0.31}_{-0.29}$ & $5.0^{+6.2}_{-4.1}$ \\
GBM MCMC & $0.06^{+0.08}_{-0.03}$ & $1.72^{+0.21}_{-0.19}$ & $3.7^{+5.0}_{-2.6}$ \\
\hline
Joint MCMC & $0.07^{+0.10}_{-0.04}$ & $1.67^{+0.26}_{-0.25}$ & $5.6^{+5.7}_{-4.7}$ \\
\hline
\end{tabular}
\tablefoot{Posterior estimates for the SFR-scaled rate $R$, PL index $\beta$, and minimum isotropic-equivalent energy $E_{\rm iso, min}$ from MCMC fits to individual and joint MGF samples. "Burns (2021) Results" are the values from the \citet{2021Burns} analysis. "Burns (2021) MCMC Benchmark" are the results from the MCMC fitting using the \cite{2021Burns} sample and posteriors. "Burns (2021) IPN MCMC" uses the \cite{2021Burns} samples with our posterior assumptions and the updated $T_{\rm{obs}}$. "IPN MCMC" incorporates additional IPN and GBM candidates GRB\,180128A and GRB\,231115A. Values are reported at the 90\% confidence level.}
\end{table*}
    
    The \citet{2021Burns} sample consists of six bursts located within 5\,Mpc, drawn from IPN detections. As previously stated, to correct for the known galaxy overdensity within this volume, the original results were scaled to the total local SFR within a 50\,Mpc sphere. In our analysis, we report volumetric rates scaled to both 50\,Mpc and 30\,Mpc volumes to enable direct comparison with prior work. However, we adopt 30\,Mpc as our preferred normalization distance, as it approximates the scale at which the local universe becomes homogeneous. Additionally, this choice ensures a more complete sampling from the galaxy catalog, as catalog completeness declines with increasing distance. Our modeling uses a galaxy catalog limited to this volume, with simulated events assigned to galaxies in proportion to their total SFR. Best-fit values from the MCMC analysis under each scaling assumption are presented in Tables~\ref{tab:mcmc_results} and \ref{tab:volumetric} .

    Our MCMC reproduction of the \citet{2021Burns} sample (e.g., MCMC Benchmark) yields posterior distributions that closely match the original study, validating the consistency of our approach. Across all three instrument-specific samples, the inferred values of $\beta$ are mutually consistent within uncertainties, lending additional confidence to the robustness of the population-level constraints. The inferred $E_{\rm iso, min}$ differs across samples due to differences in sensitivity and modeling assumptions, and the corresponding volumetric rates reflect the number of MGFs expected above the respective thresholds. Notably, the shift from a 50\,Mpc to 30\,Mpc scaling distance leads to systematically higher inferred rates. This is an expected outcome, given that the SFR measurement are imcomplete and get worse with distance. The total SFR enclosed within 30\,Mpc is smaller ($\sim1400\,M_{\odot}\, yr^{-1}$ at 30\,Mpc compared to $\sim4000\,M_{\odot}\, yr^{-1}$ at 50\,Mpc), requiring a higher intrinsic rate per unit volume to reproduce the observed number of events.
    
    The Updated IPN sample incorporates more recently confirmed MGFs and spans a longer observational baseline, resulting in improved completeness. The addition of GRB\,180128A and GRB\,231115A, along with updated model parameters, yields an inferred $\beta$ that remains consistent with previously published values and with results from the other MCMC fits. The inferred $E_{\rm iso, min}$ is a factor of four lower than in the Burns (2021) MCMC benchmark, while the corresponding SFR-scaled rate $R$ increases.
    
    The GBM candidate sample includes the majority of bursts in the modeled dataset and benefits from uniform detection conditions and a well-characterized sensitivity curve. The inferred $\beta$ remains consistent with those derived from the IPN analyses, while the lower value of $E_{\rm iso, min}$ reflects enhanced sensitivity of GBM to lower-energy events.

\renewcommand{\arraystretch}{1.2}
\begin{table*}[ht!]
\centering
\caption{Scaled Volumetric Rates}
\label{tab:volumetric}
\begin{tabular}{lccc}
\hline\hline
Sample & Scaling Distance & Volumetric Rate \\
       & (Mpc) & ($\times10^{5}\,\rm{Gpc^{-3}\,yr^{-1}}$) \\
\hline
Burns (2021) Results  & 50 & $3.8^{+4.0}_{-3.1}$ \\
Burns (2021) MCMC Benchmark & 50 & $3.9^{+4.4}_{-2.0}$ \\
Burns (2021) IPN MCMC & 50 & $5.6^{+5.9}_{-2.9}$ \\
Updated IPN MCMC & 50 & $6.6^{+6.0}_{-3.2}$ \\
\hline
Burns (2021) IPN MCMC  & 30 & $8.7^{+9.9}_{-4.9}$  \\
IPN MCMC & 30 & $11.2^{+9.9}_{-4.9}$\\
GBM MCMC & 30 & $8.0^{+9.4}_{-4.0}$ \\
\hline
Joint MCMC & 30 & $8.2^{+6.2}_{-3.2}$ \\
\hline
\end{tabular}
\tablefoot{Volumetric rate estimates derived from MCMC fits to individual and joint samples, scaled to the star formation rate within the indicated distance. The original \citet{2021Burns} sample is shown separately, while the Updated IPN and GBM samples, along with the combined joint model, highlighted in yellow, represent the results of the analysis presented in this work.}
\end{table*}

\subsection{Results from the Joint Sample}

    In addition to the individual sample fits, the Joint MCMC model yields a set of posterior distributions that incorporate all observed bursts within a unified framework. Simulated events are passed through the same population-level energy distribution as before, but evaluated under both GBM and IPN detection criteria. For each instrument, the predicted number of detections is computed independently based on the appropriate exposure, sensitivity, and completeness scaling. To account for overlapping detections, the expected number of joint detections is calculated using the intersection of the instrument-specific limiting factors: the IPN-limited galaxy subset and completeness, and the GBM-limited detection efficiency and observing time. This overlap in detections is subtracted from the combined total to avoid double-counting.

\begin{figure*}[ht!]
    \centering
    \includegraphics[width=0.9\textwidth]{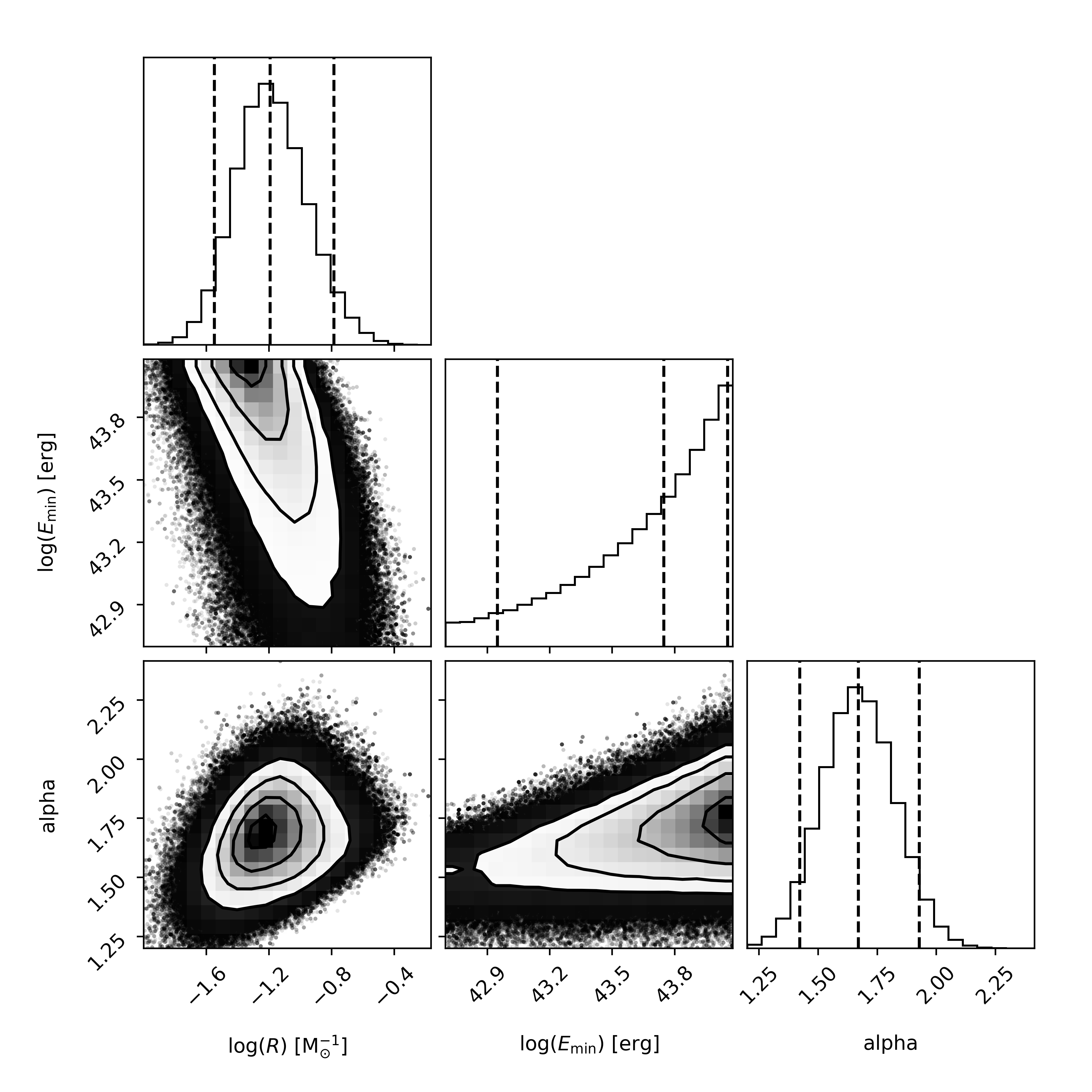}
    \caption{Posterior distributions from the Joint MCMC fit to the combined IPN and GBM samples, showing the marginalized one- and two-dimensional distributions for the model parameters: the SFR-scaled rate $\log_{10} R$, $\log_{10} E_{\rm iso,min}$, and $\alpha$. Contours in the 2D plots represent 1$\sigma$, 2$\sigma$, and 3$\sigma$ credible regions. These results represent the best-constrained model from the full joint population analysis, incorporating both spectral and detection completeness effects across instruments.}
    \label{fig:joint_corner}
\end{figure*}
    
    The Joint posterior places tighter bounds on the SFR-scaled rate $R$ and exhibits a preferred value of $\beta$ that is consistent with \citet{2021Burns} and theoretical expectations. These constraints, summarized in Table~\ref{tab:mcmc_results} and Table~\ref{tab:volumetric}, represent the most statistically robust inference in this study. The full posterior distributions are shown in Figure~\ref{fig:joint_corner}, and we adopt the Joint fit as our preferred model for subsequent discussion. 
    
    For the convenience of the reader, we also report two additional Joint-fit results under physically motivated parameter constraints. First, fixing $E_{\mathrm{iso,min}} = 1.2 \times 10^{44}$\,erg--corresponding to the lowest $E_{iso}$ observed among known MGF candidates--yields a volumetric rate of $5.5^{+4.5}_{-2.7} \times 10^5$\,Gpc$^{-3}$\,yr$^{-1}$ with $\beta = 1.76^{+0.25}_{-0.26}$. Second, fixing both $E_{\mathrm{iso,min}}$ to this value and $\beta = 1.7$, which is broadly consistent with all other fits, yields a rate estimate of $5.05^{+2.88}_{-2.09} \times 10^5$\,Gpc$^{-3}$\,yr$^{-1}$. This latter result reflects the most conservative estimate of the extragalactic MGF rate, based on physically motivated assumptions consistent with the broader set of fits.

    The similarity of inferred population parameters across the IPN, GBM, and Joint samples supports the plausibility that the candidate events originate from the same physical process. While not definitive, this consistency strengthens the case that these bursts are compatible with an MGF interpretation, given their alignment with both theoretical expectations and properties of known MGFs. Importantly, the agreement between the GBM sample, in which selection was based purely on temporal and spectral criteria, and the IPN sample with independent localizations provides strong validation for our GBM-based event selection method and the population modeling framework developed in this work.

\section{Discussion}
\label{sec:disc}

    With an observational sample and a physically motivated forward-folding modeling framework, we use the Joint MCMC model to infer key properties of the extragalactic MGF population. These results constrain the intrinsic distribution of isotropic-equivalent burst energetics, the volumetric rate of detectable events, and their connection to magnetar formation channels and recurrence behavior. Together, they offer insight into how the inferred population properties relate to progenitor demographics, observed Galactic activity, and the broader diversity of magnetar-driven high-energy transients.

\subsection{Intrinsic Energetics Distribution}
\label{sec:E_dist}

    Our MCMC analysis adopts the same PL functional form for the distribution of $E_{\mathrm{iso}}$ used in \citet{2021Burns}, parameterized by a slope $\beta$, a maximum energy cutoff $E_{\rm{iso,max}}$, and a minimum energy cutoff $E_{\mathrm{iso,min}}$. Across all three samples--Burns IPN, Updated IPN, and GBM--the inferred PL slopes are in good agreement with the $\beta \sim 1.7$ index reported by \citet{2021Burns} (see Table~\ref{tab:mcmc_results}). This consistency holds despite differences in sample size, detection thresholds, and completeness modeling. The robustness of this result across distinct instruments and modeling assumptions reinforces the conclusion that MGFs follow a moderately steep energetics distribution, in line with both theoretical expectations and prior observations of lower-energy magnetar flares. In particular, the inferred PL slope of $\beta\sim1.7$ is remarkably consistent with the predictions of starquake-driven models of magnetar activity, in which crustal strain builds up and is episodically released in discrete energy bursts, analogous to earthquakes \citep{1996Natur.382..518C}. This distribution matches the Gutenberg-Richter PL ($\beta=$1.66) seen in terrestrial seismic activity, and is reproduced in magnetar burst observations such as those from SGR\,1806--20, where fluence distributions measured over years of \integral data exhibit a similar slope \citep{Gotz_2006A&A...445..313G}. The close alignment of the observed MGF $E_{\rm iso}$ distribution with these theoretical models provides compelling support for the view that magnetic stress-induced starquakes in the NS crust drive MGFs.

    The inferred values of $E_{\mathrm{iso,min}}$ vary across samples. The GBM-only model yields the lowest threshold, $E_{\mathrm{iso,min}} \sim 3.7^{+5.0}_{-2.7} \times 10^{43}$\,erg, while the Updated IPN model returns a slightly higher value of $E_{\mathrm{iso,min}} \sim 5.0^{+6.2}_{-4.1} \times 10^{43}$\,erg. The Joint fit converges at a value of $5.6^{+5.7}_{-4.7}\times10^{43}\,\rm{erg}$. While these differences may loosely correspond to the sensitivity and sampled detection volume of each instrument, the $E_{\mathrm{iso,min}}$ posteriors are primarily shaped by the absence of lower-$E_{\mathrm{iso}}$ events within the effective volume of each survey. In all cases, the constraints favor relatively high threshold values, near the upper end of the prior range (see Figure~\ref{fig:joint_corner}), indicating that the current data sets lack sufficient spacetime volume to tightly constrain the low end of the MGF $E_{\rm iso}$ distribution.

\subsection{Temporal Structure in MGF Peak Emissions}
    
    A reasonable fraction of the identified candidates exhibit temporally complex emission. Approximately 40\% of the extragalactic sample display multiple statistically distinct pulses. This confirms that MGFs can show multi-pulsed temporal structure, a property previously documented primarily in lower-energy magnetar bursts. These findings extend the known diversity of MGF temporal behavior and demonstrate that the presence of multiple pulses is not inconsistent with a magnetar origin. While burst morphology remains a valuable discriminant in candidate selection--particularly features such as millisecond variability and hard initial spikes--it cannot, on its own, conclusively confirm or exclude an MGF origin. Instead, temporal structure must be considered alongside spectral properties, energetics, and spatial coincidence with star-forming galaxies within a comprehensive, multi-dimensional classification framework, as adopted throughout this analysis.

% \subsection{Low-mass Compact Mergers as a Magnetar Formation Channel}

%     The association of these events with star-forming galaxies, combined with the use of an SFR-weighted galaxy catalog, strongly supports a CCSN origin for their magnetar progenitors. However, low-mass compact object mergers, such as neutron star–white dwarf and white dwarf–white dwarf mergers, have been proposed as alternative pathways for forming magnetars, particularly in delayed or binary evolution channels \citep{King_2001MNRAS.320L..45K,Yoon_2007MNRAS.380..933Y,Schwab_2015MNRAS.453.1910S}. Moreover, if such mergers do produce magnetars, the high inferred MGF volumetric rate implies that each resulting magnetar would still need to generate multiple giant flares over its lifetime.

%     In addition, a low-mass merger origin is inconsistent with the host galaxy demographics of observed MGF candidates. The strong preference for star-forming hosts aligns with a CCSN origin and is difficult to reconcile with merger channels, which are expected to follow an older stellar population with delayed time distributions \citep{Fryer_1999ApJ...526..152F,Maoz_2014ARA&A..52..107M}. While low mass mergers may produce magnetars capable of flaring, the absence of MGF associations with low SFR galaxies (with the possible exception of GRB\,070201 in M31, which has a star formation rate of $0.4\,M_{\odot}\,\mathrm{yr}^{-1}$ and accounts for one out of thirteen known MGFs) suggests that such channels are unlikely to dominate the extragalactic MGF population.

\subsection{Volumetric Rate and Event Recurrence}
\label{sec:rates}

    The Joint MCMC model represents our preferred estimate of the extragalactic MGF population, incorporating both GBM and IPN detections into a unified likelihood framework. It yields a volumetric rate of $8.2^{+6.2}_{-3.2} \times 10^5$~Gpc$^{-3}$~yr$^{-1}$, scaled to the total SFR within 30\,Mpc. The rate should be understood in the context of the fitted energetics threshold ($E_{\mathrm{iso,min}}$) of the model, as determined by the posterior distribution. The reported rate is not an absolute, independent quantity, but rather reflects the occurrence of MGFs with $E_{\rm iso}$ values above the inferred $E_{\mathrm{iso,min}}$ within the modeled population. Because $E_{\mathrm{iso,min}}$ is itself a fitted parameter, the inferred rate is conditional on that threshold, making direct comparisons between models with different values nontrivial.

   When scaled by the combined star formation rate of the Milky Way and LMC ($\sim$1.95~M$_\odot$~yr$^{-1}$), the inferred SFR-scaled rate $R$ implies a Galactic occurrence rate of approximately $0.14^{+0.20}_{-0.08}\,\rm{yr^{-1}}$. Under the assumption of a stationary Poisson process, this corresponds to an expectation of Galactic events over the last 57 years of $0.14^{+0.20}_{-0.08}\,\rm{yr^{-1}} \times 57\,\rm{yr} = 8^{+11}_{-5}$. In contrast, only 3 such events have been observed in that interval. The probability of observing 3 or fewer events given this expectation is approximately 4.3\%, indicating mild ($\sim2\sigma$) tension. Our fitted $E_{\mathrm{iso,min}}$ values may reflect an incomplete sampling of the intrinsic burst energetics, particularly at the low end of the energetics distribution. Future observations are needed to determine whether the apparent overprediction results from selection effects, misclassified bursts, or a break in the underlying distribution.

    Following \citet{Tendulkar_2016}, we compare the inferred volumetric MGF rate to the expected rate of progenitor formation, which can be expressed as
\begin{equation}
    R_{\rm MGF} = R_{\rm Event} \, f_{\rm M} \, \tau_{\rm Active} \, r_{\rm MGF/M}
\end{equation}
    Here, $R_{\rm Event}$ is the rate of progenitor events capable of forming magnetars, $f_{\rm M}$ is the fraction of those events that successfully produce magnetars, $\tau_{\rm Active}$ is the timescale over which magnetars can produce giant flares, and $r_{\rm MGF/M}$ is the rate of MGFs per magnetar per unit time. We adopt $\tau_{\rm Active} \approx 10^4$~yr, consistent with the decay timescale of the magnetic field as estimated by \citet{Beniamini2019}.

    If we assume that CCSN is the only progenitor channel, and that all CCSN produce magnetars, then $R_{\rm events} = R_{\rm CCSN} = 7 \times 10^4\,\rm{Gpc^{-3}\,yr^{-1}}$ \citep{Li_2011MNRAS.412.1473L} and $f_{\rm M} = 1$. Using the lower bound found for our Joint volumetric rate ($4.9 \times 10^5\,\rm{Gpc^{-3}\,yr^{-1}}$), the average magnetar (at 90\% confidence) must produce at least $r_{\rm MGF/M} = 7$ giant flares over an average active lifetime. This represents the first empirical lower bound on the number of MGFs expected per magnetar, providing a new constraint on their recurrence behavior that is directly informed by volumetric event rates.

\subsection{Assessing MGFs as a Source of Fast Radio Bursts}
\label{MGF_FRB}

    Only one Galactic magnetar, SGR\,1935+2154, has been observed to emit an FRB-like radio burst contemporaneously with a soft X-ray event (FRB\,200428), though the energy of the flare was well below the MGF regime \citep{bochenek2020fast, Mereghetti+20}. \citet{Beniamini_2025} emphasize that while such events may arise during magnetar short bursts, there is no evidence that classical MGFs routinely generate FRBs, and many of the brightest MGFs lack any associated radio signature. Supporting this, \citet{Principe_2023A&A...675A..99P} found no gamma-ray counterparts to known FRBs in a dedicated search using \fermilat. However, one oddity in this Galactic FRB–SGR association is that the FRB itself was sub-energetic compared to the broader cosmological FRB sample. If the FRB and SGR emission energetics scale together, this suggests that more distant FRBs may arise from more energetic flares motivating the consideration of MGFs as potential FRB progenitors. This, together with the mismatch in volumetric rates, indicates that if magnetars do produce FRBs, only a rare subset are responsible.

    The volumetric rate inferred from the Joint MCMC model is notably higher than most published estimates of the FRB event rate. For instance, \citet{James_2022MNRAS.510L..18J} report a volumetric rate of FRBs above $10^{39}$\,erg of $\Phi_{39} = 8.7^{+1.7}_{-3.9} \times 10^4\,\rm{Gpc^{-3}\,yr^{-1}}$, based on CHIME detections. When compared to the MGF rate derived in this work, this suggests that at most $\sim 10^{-2}$ of MGFs could be accompanied by detectable FRB emission, assuming any are associated at all. Importantly, this should be interpreted as an upper limit on the fraction of MGFs that could produce FRBs; the true value could be substantially lower, or zero, depending on the physical mechanism and emission geometry involved.

\section{Conclusion}
\label{sec:conclude}

    Our refined search of archival \fgbm data, guided by observationally informed selection criteria, yielded four additional MGF candidates, thereby expanding the extragalactic sample and enhancing the statistical power available for population modeling. 

    The clear positive correlation observed between total isotropic energy ($E_{\rm iso}$) and spectral peak energy ($E_{\rm p}$), incorporated into the MCMC framework, enables realistic variation in spectral shape across the population and addresses a significant limitation of prior analyses, which assumed fixed spectral parameters. All three instrument-specific MCMC fits---the Burns IPN, Updated IPN, and GBM-only samples---yield consistent posterior distributions for $R$ and $\beta$ within uncertainties. Minor differences in inferred parameters are attributed to statistical variation between samples, as the modeling framework explicitly accounts for each instrument’s sensitivity and selection effects. This consistency across independent datasets reinforces the robustness of the inferred MGF population parameters.

    The Joint MCMC model represents the most comprehensive and self-consistent inference of the extragalactic MGF population to date. By combining the GBM and IPN samples under a unified likelihood function, the model accounts for distinct detection criteria, selection effects, and completeness functions for each instrument. The resulting joint posterior places tighter constraints on the volumetric rate, likely reflecting the broader spacetime volume and increased sample size accessible in the combined analysis.
    
    The inferred volumetric rate implies that MGFs are relatively common in star-forming galaxies, consistent with a CCSN origin. The rate requires that each source emit multiple flares above over its active lifetime, reinforcing recurrence as a fundamental aspect of magnetar evolution. This recurrence requirement places strong constraints on alternative progenitor channels which occur too infrequently to account for the observed population without invoking implausibly high flare rates per source.
    
    The high MGF rate also exceeds current estimates of the FRB volumetric event rate, suggesting that only a small fraction, on the order of $10^{-2}$, of MGFs produce observable FRBs. These findings imply that while MGFs may serve as progenitors for some FRBs, the radio-bright subset represents only a small fraction of the broader magnetar flare population.

    Several of the newly identified GBM candidates are sufficiently bright to warrant searches for potential additional detections to improve IPN localizations, which could help mitigate the relatively poor localization precision of GBM. Adapting our search algorithm to instruments like BATSE may yield further candidates, while extending the approach to current instruments like GBM and future observatories such as the Compton Spectrometer and Imager \citep[COSI;][]{COSI_2024icrc.confE.745T} could enable real-time classification and facilitate X-ray follow-up aimed at detecting the elusive modulated tail, the smoking-gun signature of MGFs. The results presented here demonstrate the power of unified, instrument-sensitive modeling frameworks for understanding rare astrophysical populations. With the validation of the Joint inference model and the incorporation of spectral and flux-dependent detectability, future work can now focus on refining the inferred luminosity function, constraining recurrence statistics, and clarifying the role of MGFs in high-energy transient astrophysics.

\begin{acknowledgements}
The USRA co-authors gratefully acknowledge NASA funding from cooperative agreement 80NSSC24M0035. E.N. acknowledges NASA funding through Cooperative Agreement 80NSSC22K0982. O.J.R. graciously acknowledges NASA funding support from Fermi GI large grant 80NSSC21K2038.
\end{acknowledgements}

%-------------------------------------------------------------------

\bibliographystyle{aa}
\bibliography{pop_study}

\end{document}